# Development of a multispectral stereo-camera system comparable to Hayabusa2 Optical Navigation Camera (ONC-T) for observing samples returned from asteroid (162173) Ryugu

Yuichiro Cho[a]*, Koki Yumoto[a], Yuna Yabe[a], Shoki Mori[a], Jo A. Ogura[a], Toru Yada[b], Akiko Miyazaki[b], Kasumi Yogata[b], Kentaro Hatakeda[b], Masahiro Nishimura[b], Masanao Abe[b], Tomohiro Usui[b], Seiji Sugita[a, c]

[a] Department of Earth and Planetary Science, The University of Tokyo, 7-3-1 Hongo, Bunkyo, Tokyo, 113-0033 Japan

[b] Institute of Space and Astronautical Science, Japan Aerospace Exploration Agency, 3-1-1 Yoshinodai, Chuo-ku, Sagamihara, Kanagawa, 252-5210 Japan

[c] Planetary Exploration Research Center, Chiba Institute of Technology, 2-17-1 Tsudanuma, Narashino, Chiba, 275-0016 Japan

*Corresponding author: Y. Cho (cho@eps.s.u-tokyo.ac.jp)

**Abstract**

Hayabusa2 collected 5.4 g of samples from the asteroid (162173) Ryugu and brought them back to Earth. Obtaining multiband images of these samples with the spectral bands comparable to those used for remote-sensing observations is important for characterizing the collected samples and examining how representative the sample is compared with spacecraft observations of Ryugu as a whole. In this study, we constructed a multiband microscopic camera system that enables both visual multispectral imaging at 390 (ul), 475 (b), 550 (v), 590 (Na), 700 (w), and 850 nm (x), and three-dimensional (3D) shape reconstruction of individual grain samples based on stereo imaging. The imaging system yields the images of 4096 × 2160 pixels with the pixel resolution of 1.93 μm/pix and field of view of 7.9 mm × 4.2 mm. Our validation measurements demonstrate that our multispectral imaging system, which observes the samples with the spectral bands comparable to those on the telescopic optical navigation camera (ONC-T) on Hayabusa2, yields reflectance spectra with a relative error of 3% and a 3D model with an error of 5%. These results indicate that the multiband imaging system with a 3D shape reconstruction capability yields accurate spectral and shape data on the returned samples. Using this instrument, we conducted multispectral measurements of two Ryugu samples (grains in the dishes A3 and C1) acquired from two locations on the asteroid. The average spectra of the measured Ryugu samples were flat and consistent with the global averaged spectrum of Ryugu. The 550-nm band (v-band) reflectance of the returned grains in the dishes was 2.4% on average, higher than that of the global averaged spectrum of Ryugu observed with ONC-T. This apparent difference could be because the returned grains have greater specular reflectance. In this paper, a hardware description, development, and experimental results are presented.

**Keywords**

Multispectral imaging instrument, Ryugu samples, reflectance spectroscopy, 3D shape measurements, Hayabusa2, ONC-T

# 1. Introduction

The spectrum and albedo of an asteroid provide compositional information of the asteroid surface that can be compared with those of meteorites. Absorption features in asteroid spectra are indicative of composition and can be used to infer the presence of certain silicate minerals. For example, S-type asteroids tend to have albedos >0.1 and exhibit broad absorptions due to mafic minerals at 1 µm (e.g., Pieters and McFadden, 1994). In contrast, C-type asteroids exhibit much lower albedos and flat spectrum suggestive of carbonaceous materials. Some C-type asteroids exhibit an absorption band at 0.7 µm characteristic of $Fe^{3+}$ in some phyllosilicates such as Fe-bearing serpentine (e.g., Vilas and Gaffey, 1989). Spectral slope also provides insight into the degree of space weathering, size of surface grains, and other physical properties of the surface materials (e.g., Clark et al. 2002).

The Japan Aerospace Exploration Agency (JAXA) Hayabusa2 is the first sample-return mission for a C-type asteroid (Tachibana et al. 2014) and successfully brought samples from asteroid (162173) Ryugu to Earth on December 6, 2020 (Tachibana et al. 2022). Before Hayabusa2's observations, sub-meter resolution observations of a C-type asteroid had not been realized (Sugita et al. 2019). The global observation using the telescopic optical navigation camera (ONC-T), which is equipped with multiband filters of 390 nm to 950 nm (Kameda et al. 2017; see section 2 for details). The details of the inflight calibrations of ONC-T have been given by a series of papers by Suzuki et al (2018), Tatsumi et al. (2019), and Kouyama et al. (2021). The ONC-T observations have revealed that Ryugu has a flat and dark average spectrum consistent with a Cb-type asteroid and has average reflectance factor 1.88 ± 0.17% at 550 nm, which is lower than any measured carbonaceous chondrites, suggesting that the constituents of Ryugu may be different from any sample in our meteorite collection on Earth (Sugita et al. 2019).

Remote sensing observations of the surface features, such as boulders (Michikami et al. 2019, Tanabe et al. 2021) and craters (Cho et al. 2021, Takaki et al. 2022) in the ultraviolet (UV)-to-near infrared (NIR) wavelength range revealed a variation in the color (spectral slopes) of the surface material among rather homogeneous surface spectra (Sugita et al. 2019). Such a variation suggests different degrees of solar heating and/or space weathering that the surface materials have experienced (Morota et al., 2020). The color of the samples returned from Ryugu is thus an important target for sample analyses. In addition, bright boulders of which albedos >1.5 times higher than Ryugu average were found on Ryugu. Some of them exhibit spectra consistent with those of S-type asteroids, suggestive of their exogeneous origin (Tatsumi et al., 2021). Other bright boulders show flat reflectance spectra consistent with C-type asteroids, but their variation potentially records the alteration process of the material in the parent body of Ryugu (Sugimoto et al., 2021b). The size-frequency distribution observed for those bright boulders suggests an up to % level mixing ratio for sub-millimeter grains acquired as a sample (Sugimoto et al., 2021a). Such a bright grain (albedo > 5%; Tatsumi et al. 2021), possibly embedded in a larger grain, is an important target for sample analyses.

The morphologies and roughness of the boulders are also important observables for characterizing the material of Ryugu. The boulders found on Ryugu were predominantly categorized as hummocky, mottled morphology (Type 1) or smooth angular morphology (Type 2) (Otto et al., 2021; Sugita et al., 2019). This morphological difference was associated with the boulder brightness, suggesting different lithologies and/or processes in the parent body (Sugita et al. 2019). High-resolution shape measurement of the returned sample grains would be necessary to address this aspect. These variety of spectra and morphology observations indicate that Ryugu and its parent body experienced complex evolution from the formation to the current state. Accurate understanding of such complex evolution would benefit greatly from high-precision, high-accuracy geochemical analyses of actual samples from Ryugu (Tachibana et al., 2014). Such long-awaited samples were returned from Ryugu in December 2020.

Initial measurements immediately after opening the sample container revealed the most basic properties of the returned samples (Tachibana et al. 2022). The mass of samples turned out to be 5.4 g, which is much greater than the anticipated amount. The samples collected at the first touchdown to the asteroid were stored in the sample container in Room A, while those collected at the second touchdown near an artificially formed impact crater (Arakawa et al., 2020; Honda et al., 2021) were stored in Room C. Numerous millimeter-sized grains and nearly centimeter-sized pebbles were found along with submillimeter-sized fine powder in the sample container. Materials of different colors and absorption features could have been sampled in the container. The sampling sites were on the equatorial ridge, where a spectral slope bluer than the average of Ryugu was observed; nevertheless, redder and darker materials were lofted during the first touchdown (Morota et al. 2020). Furthermore, remote reflectance measurements revealed that the 0.7-μm absorption band suggestive of phyllosilicates was generally lacking from the surface spectra of Ryugu, while a minimal amount of 0.7-μm absorption was detected along the equator (Kameda et al., 2021). Furthermore, ejecta from the artificial impact experiment performed by the small carry-on impactor (SCI) might have been collected during the second touchdown and stored in Room C. Finding and characterizing such particles, likely less-weathered materials from the subsurface of Ryugu, are very important for understanding the intrinsic properties of Ryugu's material as well as for understanding the alteration process that the sample experienced. Thus, spatially-resolved measurements of the spectrum and albedo of the numerous returned particles, and comparison with those acquired using ONC-T are important for investigating the optical properties of the sample grains, examining how representative the sample is compared with spacecraft observations of Ryugu as a whole, determining their possible origins on the asteroid, and obtaining the ground truth of the remote-sensing observations.

As a part of the curatorial efforts before distributing the samples for individual high-precision analyses, preliminary optical-, size-, and mass measurements have been conducted, which revealed a number of important properties of Ryugu sample (Yada et al. 2021). For example, using a prototype of the instrument developed in this study, averaged reflectance spectra of the samples from the rooms A and C were found flat but brighter than the average of Ryugu. Two types of surface textures (rugged or smooth) seen on Ryugu boulders were also found in the grain samples. Very low bulk density (1282 ± 231 kg/m$^3$) and high micro-porosity of 46% were reported based on the volume of grains measured by approximating each grain with an ellipsoid (Yada et al., 2021). This finding updates the macro-porosity of Ryugu estimated based on the grain density of known carbonaceous chondrites (Watanabe et al. 2019) and supports well-packed interior of Ryugu. Optical imaging showed that chondrules and calcium-aluminum-rich inclusions (CAIs) were lacking from the samples, which strongly suggest CI-like nature of the Ryugu samples.

Such preliminary optical measurements of Ryugu grains required design and construction of a new measurement system that can construct the 3D digital shape models of mm-sized grains, which are as dark as the reflectance of 2%. In addition to participating in the preliminary analyses, our instrument provides the reflectance spectra, color maps, and shape models for the curatorial catalogue of individual Ryugu grains (Database by Institute of Space and Astronautical Science (ISAS), Japan Aerospace Exploration Agency (JAXA), 2022). To study spectral and morphological variation among the grains, multispectral and microscopic imaging of grains with simultaneous shape-measurement capability was critical. A simple spectrometer or commercial-off-the-shelf (COTS) hyperspectral imager was therefore not employed.

In this paper, we describe this newly-developed measurement system equipped with the bandpass filters comparable to those of Hayabusa2 ONC-T (Kameda et al. 2016), its calibration, and detailed results of the optical measurements. In section 2, we describe instrumentation developed for multiband imaging using wavelengths comparable to the ONC-T onboard Hayabusa2, as well as 3D shape measurements of samples. In section 3, the results of the calibration experiments are illustrated to show the capability of the instrument, including accuracy, precision, and measurement uncertainties. In section 4, the reflectance spectroscopy results for the Ryugu samples are reported.

## 2. Instrument development

### 2.1 Instrument design

This section describes the overall design of our instrument, which measures the intensity of the light scattered at the sample surface for multiple wavelength bands. The reflectance of the sample was obtained by dividing the intensity of the scattered light by that of a diffuse reflectance standard.

The samples from Ryugu were placed in a glovebox installed in a class-1000 clean room in the Hayabusa2 curation facility at the Institute of Space and Astronautical Science (ISAS), Japan Aerospace Exploration Agency (JAXA) (Sagamihara, Japan). Sapphire sample dishes (20 mm in diameter) filled with the grains of Ryugu samples were placed on a manual XYZθ stage in the glovebox. Detailed sample handling procedure after opening the Earth-return capsule is described by Yada et al. (2021). A quartz cover glass on the sample dish was removed by a vacuum tweezer just before the multispectral measurements. The samples were observed through a 12-mm-thick soda-lime glass window. The density of the glass window was 2.5 g/cm$^3$. The vertical and horizontal positions of the samples were adjusted using the manual stage. The glovebox was equipped with light-emitting diodes (LEDs) along its four sides to illuminate the samples during sample handling.

The multispectral imaging system was installed on top of the glovebox with bolts. The size of the multiband imaging system was 600 × 600 × 500 mm, excluding the halogen lamp house and motor controllers that were placed at the side the glove box. Straylight was carefully removed from the multispectral imaging system: the room lights were turned off during the measurements, the windows to the hallway or next rooms were covered with blackout curtains, the whole instrument was covered by black anodized aluminum sheets to shut out any ambient light that could come from LED indicators or displays of other instruments in the room, and the lamp house was placed outside of the covered instrument box to avoid its straylight from reaching the measurement system. The reflection from the chamber glass was present as the glass was not anti-reflection coated. Nevertheless, the effect of the reflected light was negligible because such light did not reach the aperture of the lens geometrically nor illuminate the samples.

The measurement system was constructed with frames made of black anodized aluminum. The optical components and motors were mounted on the aluminum frames using posts, post holders, clamps, and other post assembly. The frames were tightly fixed to the glove box with multiple bolts. The temperature in the clean room was monitored and controlled at 23°C. The constant ambient temperature was advantageous for stabilizing the performance of the measurement system, such as the dark current of the complementary metal oxide semiconductor (CMOS) sensor.

The optical system comprised a light source that cast light of different wavelengths and a CMOS camera for sample observation. The camera was mounted on top of an automatic rotation stage to enable imaging of a sample from different angles (Fig. 1). The precision of angular positioning was ± 0.004°. The motor position was measured with an encoder equipped in the rotary stage. The measurement system was aligned such that the viewing direction and rotation axis of the camera intersected with the spotlight center on the sample surface. To obtain a 3D shape of a sample, the rotation mechanism allows observations of a grain sample from different viewpoints, ranging from nadir pointing (0°) to an emission angle of 30°. The camera was designed to observe the sample from the top (incidence (*i*), emission (*e*), phase (*α*) angles = 30°, 0°, 30°) configuration at a specific rotation angle, referred to as the "home position". To achieve this configuration, the rotational axis was tilted with respect to the vertical by 15°. The optical axis of the camera was tilted by another 15° with respect to the rotational axis. The angle of 15° was selected to maximize the parallax in the limited space above the glovebox. The rotation stage was connected to an automated z-axis stage used for fine adjustment of the focus position. A Python-based software was developed and used to automate the observations. Its functions include setting experimental parameters such as exposure time, gain, and number of images taken for each configuration, designating rotation angles for stereo analysis, initiating measurements, changing filters, controlling z-positions during image acquisition, and saving image data.

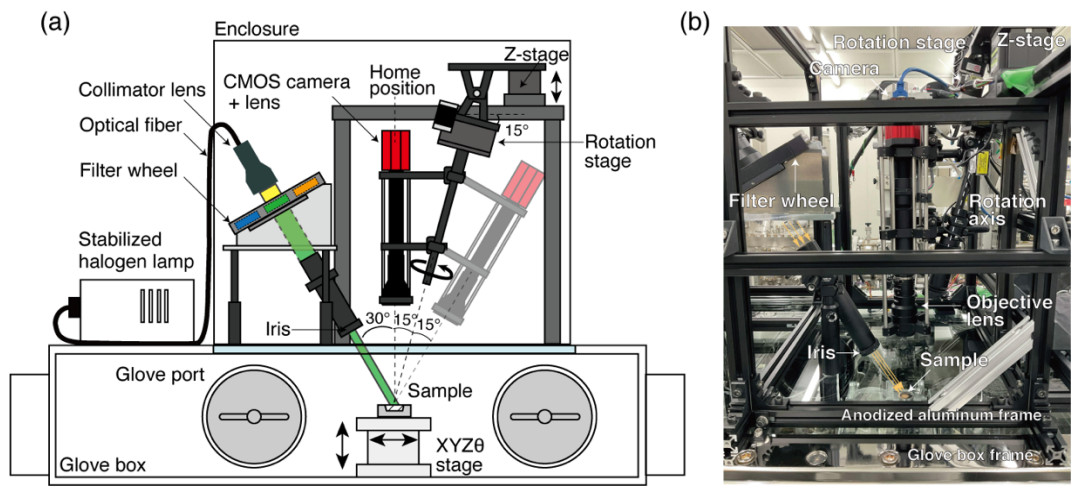

Figure 1 (a) Schematic of the instrument developed in this study. (b) Image of the instrument installed in the Hayabusa2 curation facility. The instrument was covered by black anodized aluminum sheets during the measurements.

## 2.2 Characterization of instrument components

This section provides further details of each component of the instrument and characterizes them according to the experimental setup.

*2.2.1 Light source*

The sample was illuminated by light of different colors produced via bandpass filters. A stabilized halogen lamp (Luminar Ace LA-150FBU, Hayashi Repic) was used as the light source. A cold mirror in the original lamp housing was removed to enhance the light intensity in the NIR wavelengths. A 1-m optical fiber bundle was used to connect the lamp housing with a collimator lens (C-mount lens with the focal length of 100 mm). The light from the fiber optics was collimated using a lens with the divergence angle of 0.8°. The focus position of the collimator lens was adjusted to intentionally defocus the spotlight at the sample surface, and thus eliminate the pattern of the light source, particularly that of the fiber bundle.

The intensity of the light from the halogen lamp was 0.9 mW/cm$^2$ on the glass window, which is two orders of magnitude lower than that of the solar radiation on the asteroid. The intensity of the light was designed to stabilize at a level of <1% based on a feedback control using a photon sensor assembled in the lamp unit itself. Maintaining this level of light stability during image acquisition at different bands is critical for multiband spectroscopy. Thus, we measured the temporal stability of the light using a spectrometer (HR4000, Ocean Optics) to ensure that the intensity of the incident light did not change during the sample measurements. The intensity of the spectrometer signals between 300 and 900 nm was integrated to derive the total light intensity at a certain time. In the test experiments conducted on different days, the fluctuation of the light intensity was monitored for 6 and 25 h.

Instead of installing a filter wheel on the camera head like the ONC-T, a motorized filter wheel equipped with six bandpass filters was placed at the outlet of the collimated light to select the wavelength of the incident light. Although this setup differs from ONC-T, this configuration enabled multispectral imaging analyses in the UV to NIR wavelengths equivalent to those performed by the ONC-T (Kameda et al. 2015, 2017) with a much more compact instrument that can fit in the space above the glovebox in the curation facility. Furthermore, this configuration ensured that the color of the incident light is the same over the camera FOV because the center wavelength of a filter changes if the light passes through the filter at different angles. Imaging in an intended color was therefore not assured unless collimated light passes through the filter. Six bandpass filters were selected to make the band center wavelength and bandwidth comparable to those of ONC-T. The filter wheel had eight filter mounting holes, six of which were installed with ul (band center at 388 nm), b (475 nm), v (551 nm), Na (588 nm), w (699 nm), and x (852 nm) filters. The p-band filter (950 nm) was not included in our setup because the band was not frequently used for nominal observations by ONC-T. In addition, the quantum efficiency of the CMOS camera was low in this wavelength. The other two were used for blank (no filter) and dark (light block). The diameter of each bandpass filter was 25 mm. Additionally, the actual transmittance of each filter was measured using a spectrometer. The band width and band center of each filter are compared with those on ONC-T (Table 1).

Table 1. Band centers ($\lambda_c$) and widths of the filters used in this study compared with those of ONC-T.

| Filter | This study | | ONC-T[a] | |
|---|---|---|---|---|
| | $\lambda_c$ (nm) | Bandwidth (nm) | $\lambda_c$ (nm) | Bandwidth (nm) |
| ul | 388 | 41 | 390 | 45 |
| b | 475 | 25 | 480 | 25 |
| v | 551 | 26 | 549 | 28 |
| Na | 588 | 11 | 590 | 10 |
| w | 699 | 24 | 700 | 28 |
| x | 852 | 40 | 859 | 42 |

[a]Kameda et al. (2017)

A manual iris was placed between the filter wheel and sample to control the beam diameter on the sample from 2 to 10 mm. Because the grain samples are expected to be dark (approximately 2%) and as small as 2 mm, illuminating only the samples is critical for suppressing stray light due to reflection or scattering from the wall of the sapphire glass sample holder (Yada et al. 2021).

*2.2.2 Camera system*

This section describes the experimental setup used to characterize the imaging system, the results of which are detailed in Section 3.

*Sensor and optics*

A monochromatic CMOS camera (Kiralux CS895MU, equipped with a Sony IMX267LLR-C image sensor) was used for image acquisition (Table 2). Because of the mass/size limitation of the space above the curation chamber, we did not select a camera with active cooling capabilities, at the expense of increased dark noise when exposure time is longer than 10 s. A C-mount lens ($f$ = 100 mm) was used in the camera because of its compactness. However, this C-mount lens has the shortest working distance (i.e., distance between the lens tip and the sample) of 1200 mm and did not meet the requirement of the high-resolution observation for sample analysis (<5 µm/pix). Thus, the C-mount lens was attached to the camera with an extension tube to enable the measurements at the working distance of 136 mm, which was determined by the distance between the sample and top of the chamber glass. The field of view (FOV) of the macro-lens system was 7.9 mm × 4.2 mm at the focus position, leading to the pixel resolution of 1.93 µm/pix that can resolve the 2–3-mm Ryugu samples with 1000–1500 pix. Two irises were placed at the bottom tip of the lens and immediately in front of the CMOS sensor to prevent stray light reaching the sensor.

The whole camera system was mounted on a z-stage to correct the chromatic aberration of the optical system. The z-position of the camera was set at the best focus position specific to each filter, and it was measured and determined prior to the sample analyses. This adjustment ensured the same image quality among the different bands.

A measurement of a grid distortion target of a 500-µm grid spacing showed that the lens distortion was less than 10 pixels (i.e. 20 µm) within an image. This level of distortion is negligible when building the 3D shape models of mm-sized grains.

Table 2. Specifications of the CMOS camera used in this study.

| Parameter | Value | Notes |
|---|---|---|
| Sensor type | CMOS monochrome | |
| Pixel size | 3.45 μm × 3.45 μm | |
| Pixel format | 4096 pix × 2160 pix | |
| A/D bit length | 12 bit | |
| Quantum efficiency | >70% | 525–580 nm |
| | Approx. 50% | at 400 nm |
| | 20% | at 850 nm |
| Full well | > 10,650 e- (typical) | |
| Readout noise | < 2.5 e- | |
| Focal length | 100 mm | |
| Field of view | 7.9 mm × 4.2 mm | at the focus position |
| Pixel resolution | 1.93 μm/pix | |

*Sensor linearity*

Image data are acquired at different exposure times and/or analogue gains because of different light intensities and sensor efficiencies among different bands and samples. Characterizing the behavior of the signal intensity in terms of these sensor parameters is therefore critical for accurate spectral analyses. To characterize the behavior of the signal intensity as it approached the saturation level (4,095 in digital numbers (DN)) and low level (< 100 DN), the linearity of the sensor was evaluated using a stabilized halogen lamp. We acquired images of a Labsphere 99% Spectralon sample illuminated by the halogen lamp with various exposure times of 0.036 (shortest possible), 0.1, 0.5, 1, 5, 10, 50, 100, 200, 300, 500, 700, 1000, 1300, 1500, 1700, 2000, 2500, 3000, 5000, 10 000, and 15 000 ms. Four images were acquired at each exposure time and averaged for noise reduction. The sensor gain was 0 dB in every image. Subsequently, the corresponding dark images were acquired and subtracted. Then, the DN values of a region of interest (ROI) of 200 pix × 200 pix at the center of the FOV were averaged as the intensity at a given exposure time.

*Gain and signal intensity*

The sensor gain is another parameter that controls the signal intensity. To characterize the effect of the gain with respect to DN, the images of the 99% Spectralon sample illuminated by the halogen lamp were acquired at different gain values (0, 5, 10, 15, 20, 25, and 30 dB). The exposure time was 6 s for the lower gains (0–15 dB) and 1 s for higher gains (15–30 dB) to avoid saturation of the sensor at the high gain values. Dark images were acquired with the same gain and exposure time and were subtracted from each image.

*Flat field*

The flat-field image of our camera system was obtained by averaging 30 images of another sample of 99% Spectralon in the glovebox, which was tailor-made for operation in the glovebox. The Spectralon possessed a diameter of 38 mm, which is much larger than the FOV of the camera. For each image, the position of the Spectralon was changed by several tens of pixels using the manual stage to average the surface texture of the Spectralon sample and make the target appear as uniform as possible. The illumination source comprised of white LEDs installed along the four sides of the glovebox because these lights provided the most uniform illumination source. The exposure time and gain of the camera were 5000 ms and 0 dB, respectively.

## 2.3 Multispectral analysis of calibration samples

To investigate the end-to-end performance of the multispectral camera system developed in this study, we acquired the reflectance spectra of some samples of known reflectance spectra. This section describes the samples and measurement procedures using the designed camera system. The samples were independently measured with a calibrated spectrometer for validation. The comparisons of the results are discussed in Section 3.2.

*2.3.1 Samples*

The following samples were measured using the multispectral camera system and the COTS spectrometer (Ocean Optics HR4000, the same one used for assessing the light-source stability in Section 2.2.1), a 1.5-inch Labsphere 2% Spectralon disk (black Spectralon, hereafter), a slab of the Murchison meteorite, and a slab of the Allende meteorite. The Murchison meteorite had a flat cut surface but was not polished. The Allende meteorite slab had a rough surface, formed when the meteorite was crushed by a hammer. The black Spectralon was selected to validate our instrument's capability to measure a sample as dark as Ryugu (reflectance approximately 2%) without the contribution of stray light. The carbonaceous chondrites with natural textures were selected to validate the capability of our instrument as a multiband imager, although their reflectance was higher than Ryugu's. The 99% Spectralon was measured in each reflectance measurement as a reference target.

*2.3.2 Measurement procedure*

For the multispectral analysis, the camera was set at the home position; specifically, observing the sample from the (30°, 0°, 30°) configuration. Images of the 99% Spectralon were obtained for every band as a reference.

Twenty images were acquired and averaged for each sample that was placed below the camera at the focus position. The focus was adjusted using a v-band image. The camera settings for each band (exposure time and gain) are listed in Table 3. For accurate dark subtraction, the bias level or "black level" was set to a value such that the dark frame pixels had around 100 DN. These measurements were conducted outside of the glovebox because such samples could not be placed in the clean room to avoid contamination of the Ryugu samples. Because chamber glass absorbs more UV light than visible light, the spectrum of incident light becomes redder than that without the glass. However, the different color of the incident light does not affect the reflectance of a sample because the images of the illuminated samples are divided by those of the Spectralon.

*2.3.3 Data analysis*

The images of the samples were processed using a data-reduction pipeline comparable to that used for the images obtained by ONC-T (Fig. 2). Raw image(s) were acquired for the same sample in the same bandpass filter. In the case of multiple images, the images were averaged; pixels showing the values higher than 3300 DN (discussed in section 3) are identified; the dark images were subtracted. Moreover, the signal intensity was scaled if measurements were conducted at different gain values, and it was normalized with respect to exposure time. This level image corresponds to "L2a," as defined for the images acquired using ONC-T. The high-intensity pixels identified earlier were then replaced with those acquired with short-exposure images on a pixel-by-pixel basis for a high-dynamic-range (HDR) measurement. The "short" exposure time was typically 1/7 of the longer exposure time, but was selected depending on the samples. The image was then divided by the flat-field image to correct the sensitivity loss due to the vignetting of the optics (L2b). The image was divided by that of the 99% Spectralon of the corresponding band to obtain a reflectance image in each band (e.g., v-band reflectance map). Note that flatfielding does not affect reflectance maps because sample and Spectralon have the identical flatfield.

Then, spatial co-registration is performed by a pattern-matching algorithm because the position and magnification of each image could slightly differ when imaged at different bands. After co-registration, the multiband spectrum of a sample (i.e., reflectance vs. wavelength) can be obtained by deriving the reflectance in the same ROI for the multiple bands. The color map of a sample, such as a *b/x* ratio map, can be obtained by dividing the sample's b-band image by its x-band image.

Table 2. Measurement conditions for validation samples.

| Black Spectralon | Murchison meteorite | Allende meteorite | 99% Spectralon |
|---|---|---|---|

| Filter | Exposure time (ms) | Gain (dB) | Exposure time (ms) | Gain (dB) | Exposure time (ms) | Gain (dB) | Exposure time (ms) | Gain (dB) |
|---|---|---|---|---|---|---|---|---|
| ul | 4,000 | 30 | 6,000 | 20 | 2,000 | 25 | 2,500 | 10 |
| b | 1,000 | 25 | 2,400 | 10 | 1,200 | 10 | 800 | 0 |
| v | 500 | 20 | 1,600 | 5 | 600 | 5 | 400 | 0 |
| Na | 1,000 | 20 | 1,600 | 10 | 1,200 | 5 | 800 | 0 |
| w | 500 | 20 | 800 | 10 | 600 | 5 | 300 | 0 |
| x | 2,000 | 25 | 3,200 | 15 | 1,200 | 15 | 2,000 | 0 |

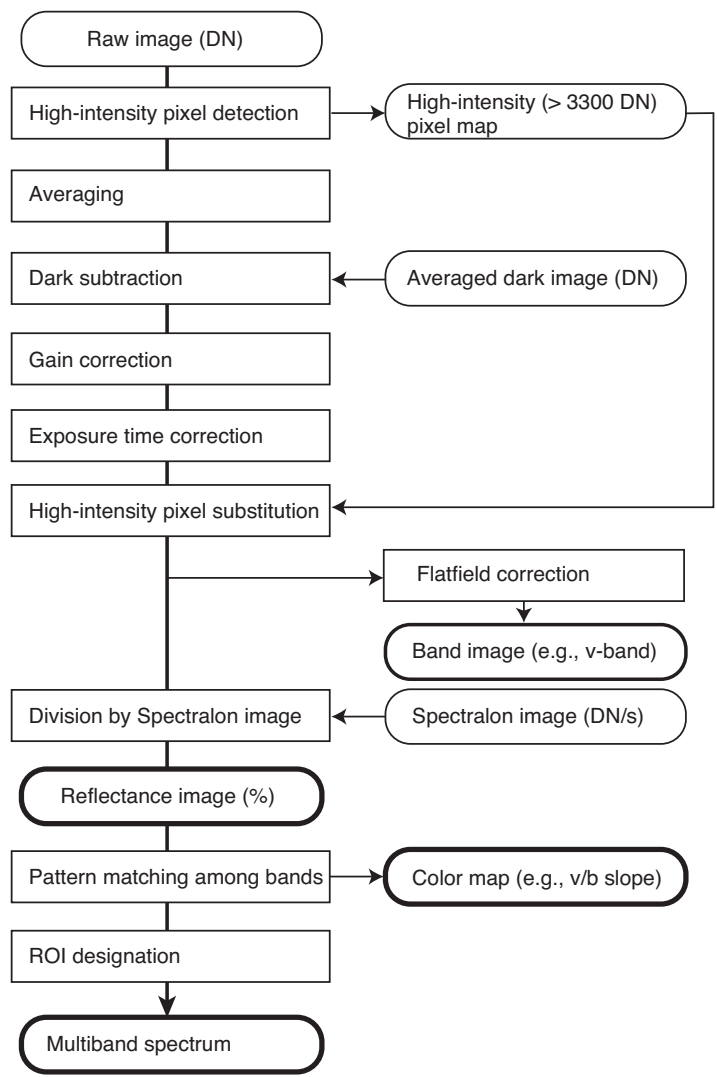

Figure 2. Data processing pipeline. The steps are as follows: (1) raw image, (2) mapping pixels having > 3300 DN for HDR pixel substitution, (3) averaging, (4) dark subtraction, (5) gain correction, (6) exposure time correction, (7) high-dynamic range pixel substitution, (8) flat field correction for individual band images. (9) Spectralon division to obtain reflectance map. (10) Multiband spectrum and color map are obtained after co-registration among different reflectance images.

*2.3.2 Spectrometer setup*

To assess the instrument's multispectral observation capability, the reflectance spectra of the samples were obtained through the HR4000 spectrometer independently from the measurements obtained using the multiband camera system. The continuous spectra covering from 350 to 900 nm at the spectral resolution of 0.25 nm/pix were compared with those obtained with the multispectral imaging system developed in this study.

To obtain the continuous reflectance spectra, the stabilized halogen lamp was used to illuminate the samples. Because the light intensities in the UV (< 450 nm) and NIR (> 750 nm) ranges were much lower than that in the visible (VIS) range, the spectra were recorded for UV, VIS, and NIR ranges under different optimized conditions. In the VIS wavelength range, the 99% and 2% Spectralon disks were measured with exposure times of 100 and 5000 ms for 100 averaging, respectively. To obtain high signal-to-noise ratio (S/N) spectra in the UV and NIR ranges, short- and long-pass filters were placed in front of the spectrometer's inlet. These filters allow the exposure time to be extended without saturating the detector in the VIS range where the light is much more intense; this could cause smearing of the pixels for UV or NIR wavelengths, yielding inaccurate reflectance values in these ranges. The extended exposure times were 700 ms (>750 nm) and 1000 ms (<450 nm) for the 99% Spectralon, and 50 000 ms (for both >750 nm and <450 nm) for the 2% Spectralon. The dark spectra were recorded when the light was turned off and subtracted. The 100 and 50 spectra were averaged for the 99% and 2% Spectralons, respectively. The average spectra in the spectrometer's FOV (approximately 5 mm in diameter) were obtained from the (30°, 0°, 30°) configuration. The FOV of the spectrometer was targeted at that of the multispectral camera to obtain the spectra from the same area. The results of these measurements are shown in Section 3.2. Although we used the COTS spectrometer for the verification of our instrument, such continuous spectra were not obtained for the grain samples from Ryugu. This is because the FOV of the spectrometer was much larger than the individual grains and acquiring accurate spectra from such dark small samples was not feasible in our setup.

## 2.4 Stereo measurements

Our instrument was designed to conduct stereo imaging of the samples at multiple angles to obtain 3D digital elevation models (DEMs) of the Ryugu samples. Such data will be used for morphological or photometric analyses. This section describes the test sample, experimental setup, and analytical procedure for the stereo imaging.

*2.4.1 Sample*

A fragment of natural graphite with the size of approximately 3 mm × 2 mm × 1 mm was imaged for testing. This sample was selected because of its relatively low albedo and the presence of reflective facets that resemble some samples from Ryugu (Yada et al. 2021). The sample was placed on an aluminum dish for observations.

*2.4.2 Measurement and analysis*

Images of the sample were acquired from multiple viewing angles by rotating the camera head on the automatic rotation stage (Fig. 1). The stage was attached to the frame but was tilted from the vertical direction by 15° (Fig. 1). The stage allowed rotation of the camera around the axis in the azimuth angles from −180 to +70°, where 0° refers to the home position used for multispectral analysis. The angle range was limited by spatial constraints, such as other instruments installed on the glovebox, as well as the frames supporting our camera system. The angular rotation rate was 5 °/s and was controlled using software. Eighty images were acquired at intervals of 3°.

Unlike the multispectral measurements where samples are illuminated from the incidence angle of 30°, we obtained images of the samples illuminated by surrounding white LED lights from multiple directions. This configuration enabled image acquisition of the samples without shadows. This was practically important because shadowed areas do not yield the information of the shape. The test measurement using the graphite was performed in advance before the system was installed in the curation facility.

A software tool based on the structure-from-motion modelling method (Agisoft Metashape) was used to obtain the DEM of the sample from the image sequence. The DEM was modelled as a polyhedral object. The length scale of each triangular polygon was typically 10 μm. The horizontal (XY) scale was calibrated with the standard scale target of 500-μm meshes. The vertical scale (grain height) was calculated using the focal length of the camera optics. The same software was also used for the shape modelling of the boulders of Ryugu (Sakatani et al., 2021). The uncertainty of the shape model measurements was evaluated by repeating the shape model construction of a graphite grain. Details are described in section 3.3.

## 2.5 Measurements of samples from Ryugu

Finally, we measured two samples from Ryugu using the developed instrument. The results of the multispectral analysis are reported in section 3. The shape modeling of individual Ryugu grains via stereo measurements is reported in a separate work (Yabe et al. 2022).

*2.5.1 Sample*

The samples were retrieved from Rooms A and C of the sample container and filled in sample dishes A3 and C1 (Yada et al. 2021). The Ryugu samples in the holder varied in size from millimeter-sized grains to finer particles of tens of micrometers. The inner diameter of each sample dish was 15 mm.

*2.5.2 Measurement procedure*

The reflectance spectra of the samples were measured following the procedure described in section 2.3. An FOV of 8 mm × 4 mm was imaged on the sample. The diameter of the spotlight was adjusted to approximately 10 mm with the manual iris (Fig. 1) to ensure that the incident light did not directly illuminate the sample holder. The image was divided into 12 ROIs of 600 pix × 600 pix at an approximate center of the FOV (each ROI was approximately 1.2 mm × 1.2 mm). To derive the averaged spectra of these two dishes, we obtained the average spectra in each ROI. One part of the error in the average spectrum of each dish was estimated based on the bootstrap method: the average spectra for 6 ROIs were calculated for $_{12}C_6 = 924$ combinations. Then, the overall measurement errors were calculated as a root sum square of the standard deviation of the average spectra calculated above and the uncertainty of our measurements, as estimated in section 3.2.1. The exposure time and gain of individual measurements are summarized in Table 4. Image acquisition was conducted with two sets of exposure times to enable HDR measurements. Short-exposure images were acquired with the exposure times 1/7 of the long-exposure images shown in Table 4. The pixels exceeding 3300 DN in the long-exposure images were replaced by those in the short-exposure images after normalizing the intensity with exposure time. Saturated pixels were excluded from reflectance measurements. Shadows were not excluded from the measurements to compare the spectra with those obtained from Ryugu, in which no shadow removal was applied.

Table 4 Measurement setting for multiband spectroscopy of Ryugu samples. Short exposure images were acquired with the exposure times 1/7 of those shown here.

|  | A3 |  | C1 |  | 99% Spectralon |  |
|---|---|---|---|---|---|---|
| Filter | Exposure time (ms) | Gain (dB) | Exposure time (ms) | Gain (dB) | Exposure time (ms) | Gain (dB) |
| ul | 14,800 | 20 | 14,800 | 20 | 12,700 | 0 |
| b | 17,400 | 0 | 17,400 | 0 | 1,500 | 0 |
| v | 7,000 | 0 | 7,000 | 0 | 600 | 0 |
| Na | 13,100 | 0 | 13,100 | 0 | 1,100 | 0 |
| w | 7,000 | 0 | 7,000 | 0 | 600 | 0 |
| x | 12,400 | 15 | 12,400 | 15 | 6,000 | 0 |

## 3. Results and discussion

### 3.1 Calibration and characterization of instrument

In this section, we report the results of the experiments for characterization of the instrument.

*3.1.1 Light source*

The measured fluctuation of the total intensity from the halogen lamp was −0.5% to +0.5% peak-to-peak over 1–25 h (Fig. 3). Our results indicate that the reflectance spectroscopy measurements should start >1 h after turning on the lamp to ensure that the light intensity was stable during the measurements. The error in the reflectance measurements due to the variation in the incident light intensity was therefore estimated to be <1% because the interval of a Spectralon measurement and sample measurement is no longer than 10 h.

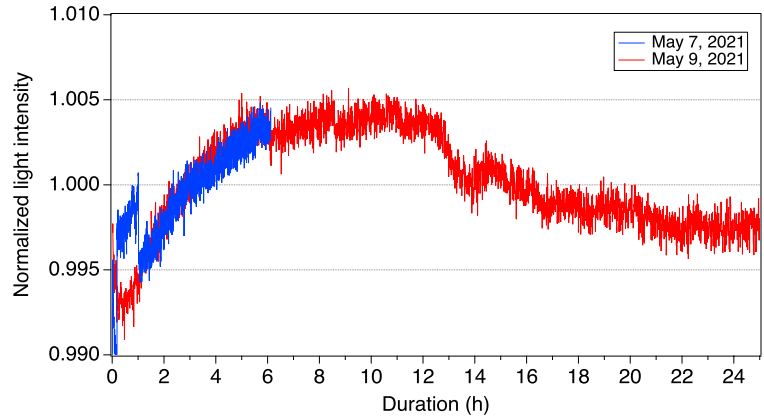

Figure 3. Temporal stability of the light intensity of the halogen lamp. The lamp was turned on at 0 h. The variation in the light intensity was less than ± 0.5% over several hours after some fluctuation in the first hour. The light intensity is normalized by the average after 0.5 h.

*3.1.2 Camera system*

*Flat field*

Figure 4 shows the obtained flat field of the camera system. The sensitivity of the system was constant at 3% in the central 2000 pix (4 mm) of the horizontal FOV. The system exhibited vignetting (down to 50% compared with the center of the FOV) at the outermost 500 pix (Fig. 4). The sensitivity dropped to approximately 30% at the four corners of the image. Note that the Spectralon is much larger (38 mm in diameter) than the FOV of the camera (8 mm in the horizontal direction); thus, the edge of the Spectralon does not define the edge of the flatfield.

      A small-scale fluctuation of approximately ± 1% over 10 pix was overlaid on the larger-scale variation (approximately >1,000 pix). The small-scale fluctuation is attributed to the heterogeneity of

the Spectralon target and/or pixel-by-pixel sensitivities of the sensor, while the large-scale one is due to the optics and can be removed by the flatfield correction. Thus, the small-scale fluctuation of ± 1% would be introduced to the flat-corrected images. The reflectance maps, however, do not suffer from uncertainty because the samples and Spectralon have the flatfield in common.

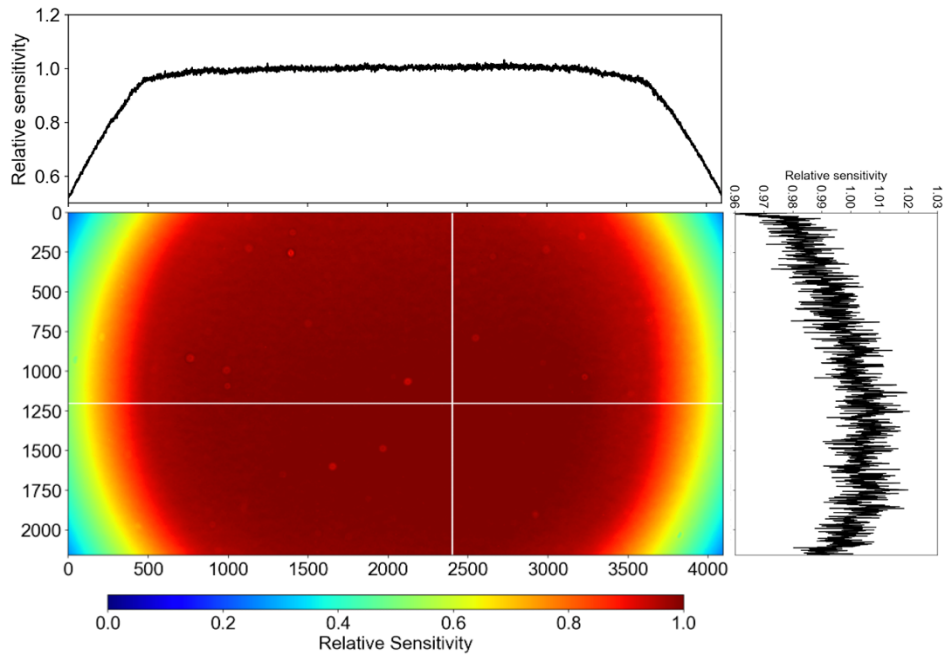

Figure 4. Flat pattern of the multiband camera system. Horizontal and vertical profiles at $x$ = 2400 pix and $y$ = 1200 pix, respectively. Note that the different sensitivity ranges are shown as profiles.

*Linearity of the CMOS detector output*

Figure 5 illustrates the relationship between the signal intensity and integration time. The signal intensity follows a linear trend until it reaches approximately 4000 DN. The linear fit accounts for the intensity between 200 and 3700 DN with a deviation less than ±1% (Fig. 5b). The deviation from the linear trend was ± 0.5% for 500–3500 DN. The signal intensity starts to deviate from the linear regression by 3% at 3700 DN, and above 4000 DN is no longer accurate because of pixel saturation. This wide-range linearity of the CMOS sensor output validates a simple linear assumption when calculating the reflectance using signal intensity, if DN values are in the range stated above.

The linearity of the signal further validates the normalization of signal intensity with exposure time, when different bands are measured with different exposure times. The uncertainty introduced through integration time correction is ± 0.5% in the pixels where the intensity is 500–3,500 DN and will slightly increase where the pixel values are 200–3700 DN. The pixels showing the intensity higher than 4000 DN should not be used or should be excluded from the analyses because these pixels suffer from saturation. We use 3300 DN as the threshold for HDR pixel substitution.

Furthermore, the signal intensity was investigated as a function of sensor gain. The intensity was exponential to the gain. An exponential regression analysis indicated that the intensity of the signals acquired with different gain values can be corrected using the following equation:

$$I_G = I_0 \exp\{(0.1148 \pm 0.0001)\, gain\} = I_0\, 10^{0.0499\, gain},$$

where $I_G$ and $I_0$ are DN values at a certain gain value and 0 dB, respectively. The uncertainty in signal intensity due to the gain conversion (deviation from the exponential trend measured at each gain value) was found to be < 1%.

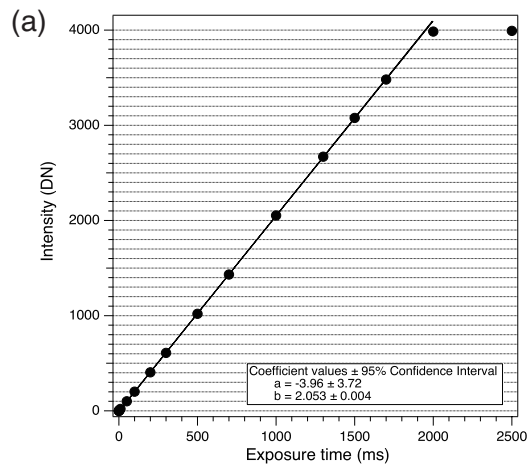

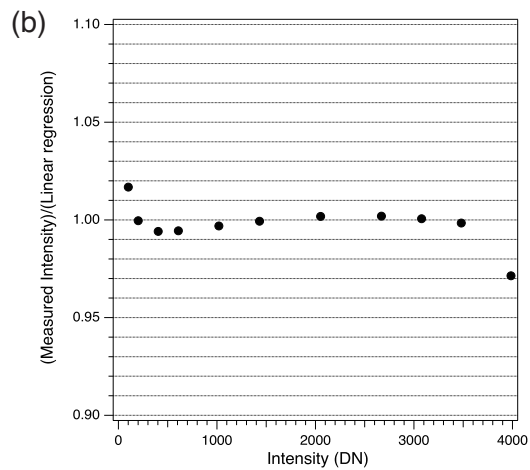

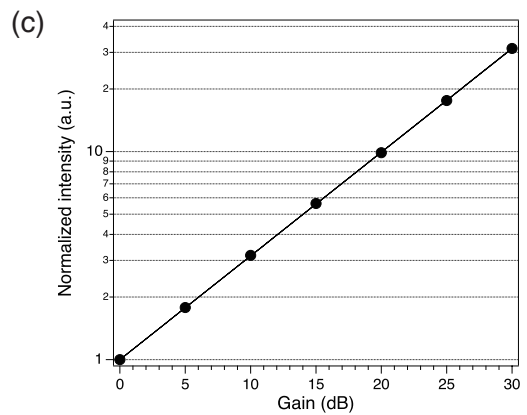

Figure 5. Linearity of signal intensity of the CMOS sensor. (a) Signal intensity (DN) vs. exposure time. (b) Deviation from the linear regression for the exposure time correction, as a function of signal intensity. (c) Signal intensity trend for different gain values.

## 3.2 Reflectance spectra of calibration samples

This section reports a comparison of multiband spectroscopy and continuous spectra to show the end-to-end performance of the instrument.

*3.2.1 Calibration with black Spectralon*

Figure 6 shows the reflectance spectrum of the black Spectralon obtained with the multispectral camera and spectrometer. The vertical error bars represent the standard deviation over nine ROIs of 200 pix × 200 pix. The multiband reflectance spectrum obtained from the multispectral images was consistent with that obtained using the spectrometer. The deviation of the multiband spectrum from the continuous spectrum (i.e., accuracy of the multiband spectrum) was 2% (e.g., 0.02% for 1% reflectance) in terms of absolute reflectance (Fig. 6). The precision of the reflectance measurement, or relative standard deviation of reflectance at each band, was 3%. This uncertainty level would be the mix of measurement error of the instrument and the heterogeneity of the black Spectralon surface. Note that a certified 2% Spectralon shows the reflectance of 2% at 2000 nm, but has the reflectance of approximately 1% in the UV-NIR range.

This result validates that the multispectral camera system developed in this study produces accurate and precise spectra to the level of 2%–3%, without any systematic error, such as that by straylight. The band ratio spectra, typically normalized by v-band reflectance, would have a relative error up to 5% as a result of error propagation.

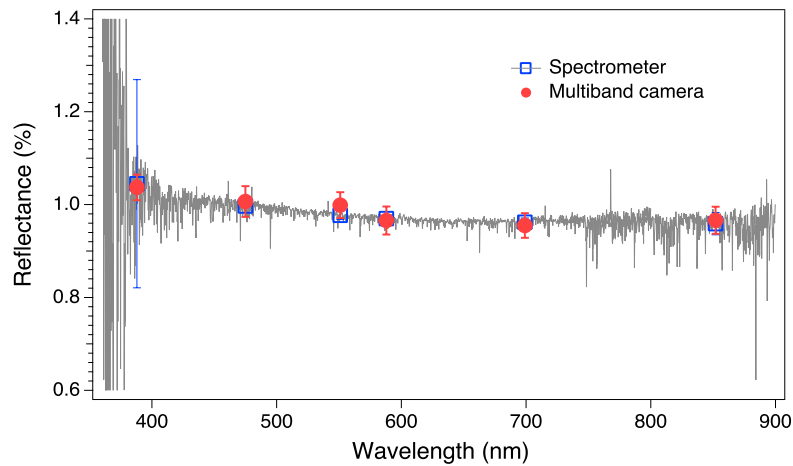

Figure 6. Multiband spectrum of the black Spectralon overlaid on its continuous reflectance spectrum obtained with a spectrometer. Vertical error bars on the red filled circles show the standard deviation of nine regions of interest. Blue open squares simulate the camera measurements by calculating the integration of spectrometer reflectance over each band width. The larger error in the continuous spectrum at < 400 nm is due to low efficiency of the spectrometer in this wavelength range.

*3.2.2 Multiband spectra of carbonaceous chondrites*

Figure 7 displays the v-band images, *v/b* slope maps, and reflectance spectra of the Murchison and Allende slab samples. The v-band images clearly show the texture of the chondrites, such as bright chondrules and CAIs embedded in the darker matrix. The slope map reveals the fine structure that coincides with the texture of the meteorites (Figs. 7c, 7d). The bright chondrules and CAIs tend to have higher *v/b* ratios than the matrix. The multiband spectra in Fig. 7e show the averages of different ROIs on the meteorite. The vertical error bars associated with the multiband spectra illustrate the heterogeneity of the natural chondrites, rather than our measurement error as estimated in the previous section. The multiband spectra are consistent with the continuous spectra obtained through the spectrometer. These results demonstrate that the multispectral camera system is capable of the multiband spectroscopy of natural rock samples, strongly supporting the readiness for analyzing Ryugu samples.

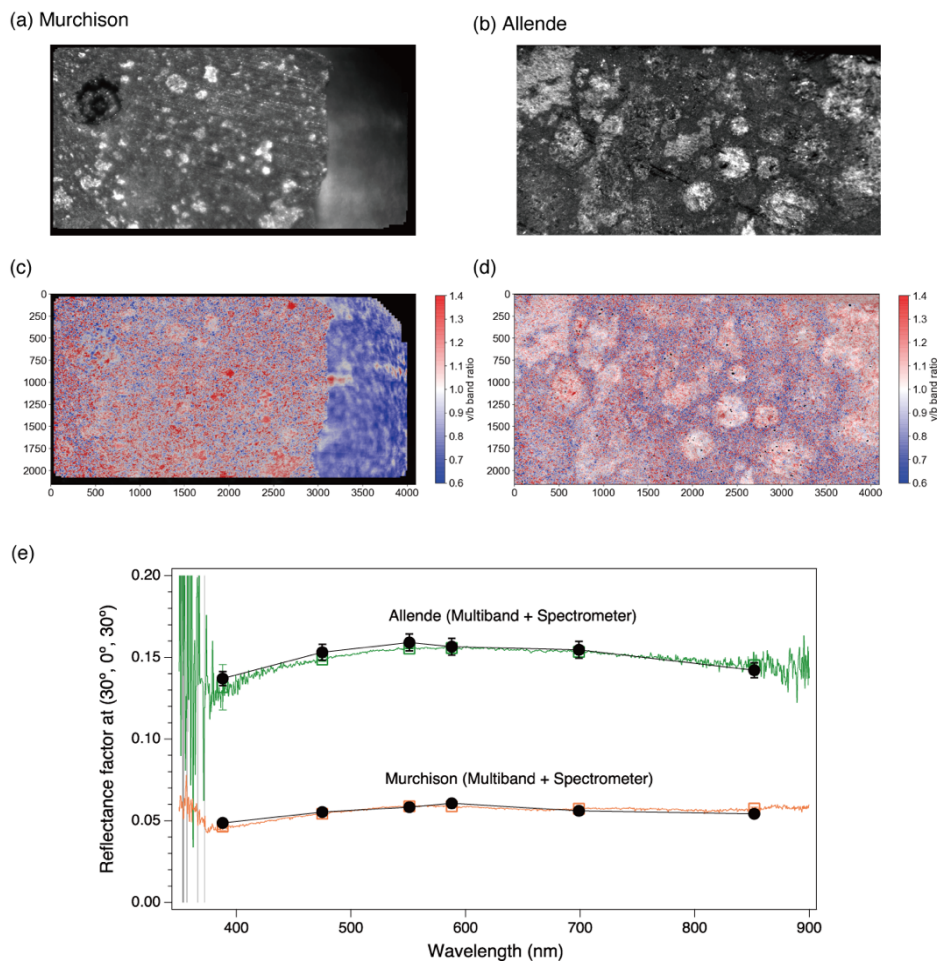

Figure 7. v-band image of the meteorite samples with ROIs. (a) Murchison and (b) Allende. Corresponding *v/b* band-ratio maps; (c) Murchison and (d) Allende. (e) Multiband reflectance spectra of the chondrites. Black filled circles show the multiband spectrum obtained with the multispectral

imaging system overlaid on the continuous reflectance spectrum independently obtained with a spectrometer. Green and orange open squares show simulated multiband spectra calculated from the continuous spectrometer data.

**3.3 Shape reconstruction**

Morphological analyses of individual returned grains would provide basic information for understanding the processes that determine the shape of the boulders/grains on Ryugu, when combined with other geochemical analyses. Accurate photometric correction can be performed using the shape of a grain (i.e., size and direction of each facet). Accurate volume measurement of individual grain would also be beneficial for density assessment of each grain.

Two DEMs of the identical graphite grain are shown in Figs. 8a and 8b. The consistency of the two DEMs, which were derived from independent measurements of the same grain from different azimuthal orientations, was evaluated by calculating differences in topography between the two models (Fig. 8c). The relative error in elevation modelled at each facet at $(x, y)$ is defined by

$$\text{Error}(x, y) = \text{abs}\{z_1(x, y) - z_2(x, y)\}/(\max(z_1) - \min(z_1)) \times 100\ (\%),$$

where $z_1(x, y)$ and $z_2(x, y)$ are the elevation of a facet at a position $(x, y)$ in the DEMs 1 and 2, respectively. $\{\max(z_1) - \min(z_1)\}$, or the height (thickness) of the graphite grain, was 1.1 mm, consistent with an independent physical measurement of the grain thickness. The root-mean-square error of the elevation of the shape model was 4.9%; the error was <5% for 78% of the total surface of the DEM. The median deviation of the elevation between the two models, calculated by Median$\{\text{Error}(x, y)\}$, was 2.4% over the entire elevation of the grain. The average area of a polygon was 43.5 μm$^2$ (length scale of 9 μm). The shape model tends to be less accurate for facets located in shadowed or saturated pixels because the number of features that can be matched was smaller in these areas (top left and bottom left areas in Fig. 8c). Furthermore, the modelling error is larger for the facets running parallel to the viewing angle of the camera because the number of stereo pairs is smaller in those areas. Although a more systematic error estimation using grains of different sizes, shapes, and albedos would be beneficial for further morphological analyses, our initial test indicates that the dark grain of a size comparable to Ryugu samples can be measured with the error of <5%. A separate paper describes the shape reconstruction of individual grains in more detail (Yabe et al. 2022).

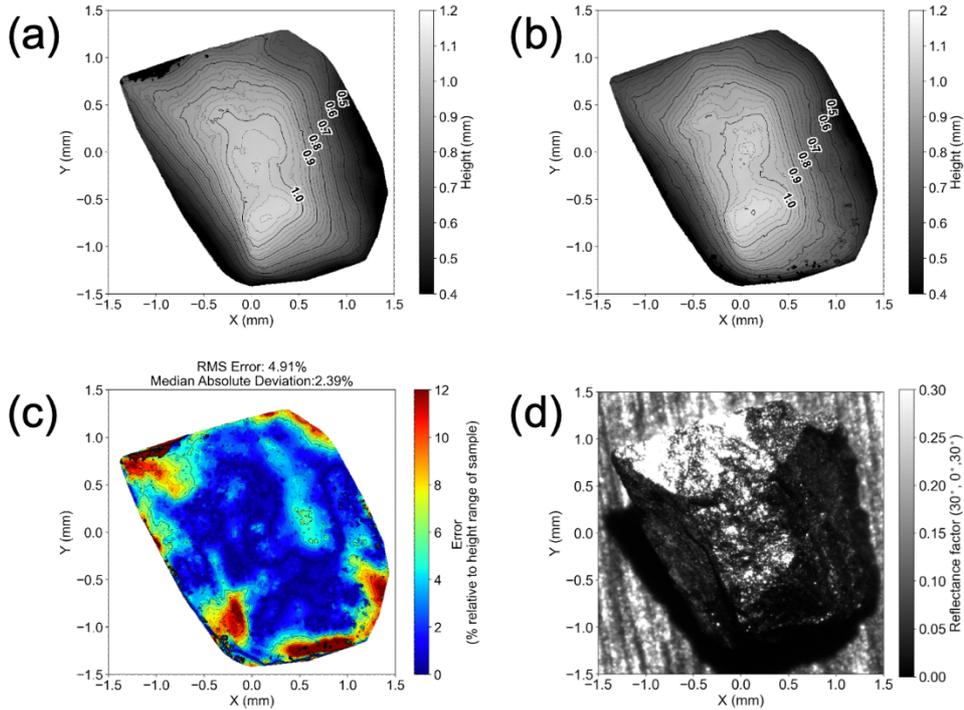

Figure 8. (a), (b) Digital elevation models of the graphite samples constructed from the same graphite grain but placed in different orientations. (c) Consistency of 3D models of the grain. (d) v-band image of the measured graphite.

## 4. Ryugu sample measurements

The v-band reflectance images of dishes A3 and C1 (Figs. 9a and 9b) show that the measured FOV of the dish A3 contains grains of 1 mm to approximately 100 μm, while C1 contains those of 1-mm to tens of microns. The samples from Ryugu show angular and relatively flat surfaces at this scale (FOV of 7.9 mm × 4.2 mm). Comparison of the Ryugu sample with Murchison and Allende meteorites (Fig 7ab) clarifies the lack of bright chondrules or CAIs in the material of Ryugu. The reflectance maps show bright facets in both dishes primarily on the larger grains. Many of these facets appear to coincide with the flat surface toward the camera, while they are rare on apparently rougher surfaces or at the side of individual grains. These observations suggest the reflective surface of Ryugu's material. Investigation of the bright boulders within the Ryugu samples would require a discrimination of the photometric effect and intrinsically bright material.

The reflectance spectra of samples A3 and C1 exhibit flat spectra with the v-band reflectance of 2.4% as an average of the measured ROIs (Fig. 9c). No systematic difference was detected between the A3 and C1 samples in terms of the bulk average spectra. Obvious absorption at 0.7 μm was not detected from the bulk spectrum. The flat spectra of the returned samples are brighter than the averaged reflectance of Ryugu measured with ONC-T (Sugita et al. 2019; Tatsumi et al. 2020). Straylight artifacts due to the scattering from the sample dish rim cannot be the reason because the incident light hit only Ryugu samples. The straylight due to the chamber glass would not be the reason either because the bright spectra of the samples are derived from μm-scale bright facets rather than uniformly enhanced brightness under a larger-scale illumination by the straylight. The reason for the different reflectance between the returned sample and global observation would be the focus of future geochemical and mineralogical analyses, but it could result from the spatially resolvable bright scattering from some of the sample facets and/or reflective opaque minerals. Another interpretation would be that the optical surface of Ryugu may have higher porosity that could cause it to be darker. The overall reflectance was still lower than other carbonaceous chondrites.

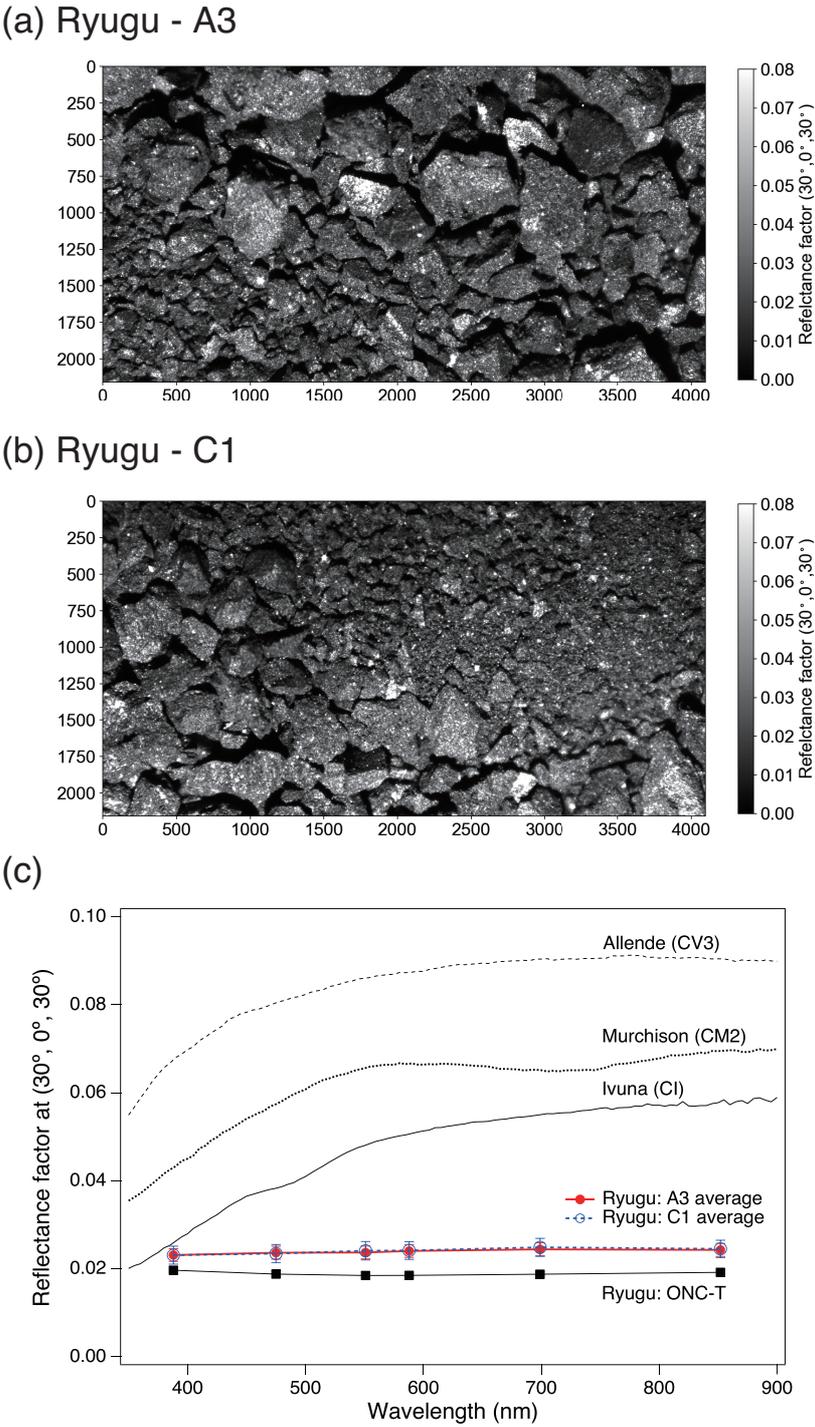

Figure 9. v-band reflectance images of Ryugu samples: (a) dish A3 and (b) dish C1. The size of the FOV is 7.9 mm × 4.2 mm. (c) Reflectance spectra of Ryugu samples compared with ONC-T and other

carbonaceous chondrites (ONC-T: Tatsumi et al. 2020; chondrites: the Brown University Reflectance Experiment Laboratory (RELAB), Pieters 1983)

## 5. Conclusions

A compact multispectral imaging system was developed to measure the samples returned from the asteroid Ryugu. The imaging system was designed to obtain multiband spectra and hyperspectral images in the bands comparable to those used with ONC-T onboard Hayabusa2, enabling direct comparison between the remote-sensing data and returned samples. The multispectral imaging system is used for curation catalogue of individual Ryugu grains as well. The camera system can acquire images of the samples at the pixel resolution of 1.93 μm/pix and FOV of 7.9 mm × 4.2 mm. The system was also designed to perform stereo imaging of the samples for morphological analyses of the grains of Ryugu.

Our calibration measurements for characterizing the instrument revealed the high precision of the components. Examples include the light source temporal stability of ± 0.5% and sensor linearity of ± 1% over a wide range of signal output. The multispectral analysis of a reflectance standard (2% Spectralon) indicates the end-to-end capability of our instrument, exhibiting accuracy and precision better than 3%. Validation measurements using the Murchison and Allende meteorites revealed that our instrument yields the multiband spectra consistent with those obtained using a conventional spectrometer, indicating the readiness for analyzing Ryugu samples.

The capability of stereo imaging was tested using a piece of natural graphite. Our characterization experiments showed the uncertainty of the DEM of less than 5%. These data validate the precise measurements of the grain morphology of mm-sized natural samples with dark but reflective surfaces.

Then, the bulk samples of Ryugu acquired from different touchdown sites were measured using the multispectral instrument developed in this study. Our data show average reflectance spectra of consistently dark (2.4% v-band reflectance) and flat shapes for both samples. These samples yielded spectra higher than the overall average of Ryugu, potentially because of the bright, reflective surface of the grains.

**Author statement**

**Yuichiro Cho**: Conceptualization, Methodology, Formal analysis, Investigation, Writing -Original Draft, Visualization. **Koki Yumoto**: Methodology, Software, Formal analysis, Investigation, Visualization, Data curation. **Yuna Yabe**: Methodology, Validation, Investigation. **Shoki Mori**: Methodology, Software, Investigation. **Jo A. Ogura**: Validation, Formal analysis. **Toru Yada**: Investigation, Resources, Project administration. **Akiko Miyazaki**: Investigation, Resources. **Kasumi Yogata**: Investigation, Resources. **Kentaro Hatakeda**: Investigation, Resources. **Masahiro Nishimura**: Investigation, Resources. **Masanao Abe**: Project administration. **Tomohiro Usui**: Project administration, Funding acquisition. **Seiji Sugita**: Conceptualization, Methodology, Investigation, Writing – Review & Editing, Supervision, Project administration, Funding acquisition.

**Declaration of competing interest**

The authors declare that they have no known competing financial interests or personal relationships that could have appeared to influence the work reported in this paper.

**Acknowledgments**

The authors thank two anonymous reviewers for their constructive and insightful comments. This study was supported by the Japanese Society for Promotion of Science (JSPS) (Grant Numbers 19K14778, 20H04607) and the JSPS Core-to-Core program "International Network of Planetary Sciences." K.Y acknowledges funding from JSPS Fellowship (Grant number JP21J20894).